\input{aipcheck}

\documentclass[
    ,final            
  ]
  {aipproc}

\layoutstyle{6x9}

\begin{document}

\title{	Signatures of (Un)particles from a Hidden Sector 
in multiparticle dynamics at Tevatron/LHC}


\author{Miguel-Angel Sanchis-Lozano}{
  address={Departamento de F\'{\i}sica Te\'orica and IFIC, 
University of Valencia-CSIC \\
 Dr. Moliner, 50 46100 Burjassot, Spain}
}

\begin{abstract}
The study of multiparticle dynamics in hadron-hadron 
collisions at Tevatron and LHC
could provide useful information on new physics in addition to the
expected signatures on the transverse plane. We suggest that 
an analysis of inclusive correlations between emitted
particles in $pp$ inelastic collisions, 
and factorial moments of multiplicity distributions,  
may be helpful in uncovering (un)particles from 
Hidden Sectors, using underlying events 
tagged by hard products like high-$p_T$ leptons and photons, and applying
stringent selection criteria like event shape variables, etc. 
\end{abstract}

\maketitle


\section{Introduction}

Multiparticle hadroproduction 
has been extremely useful along decades
to understand the strong interaction dynamics \cite{Kittel:2005fu}. 
In Ref.\cite{SanchisLozano:2008te} we explored the possibility of
applying well-known techniques based on inclusive correlations and 
moments of multiplicity distributions to the quest for
a Hidden Sector (HS) at Tevatron/LHC experiments. In all
fairness, multiparticle production
is still not fully understood, while our proposal is based on
possible departures of $\lq\lq$anomalous''  
from $\lq\lq$standard'' events. Nonetheless, one might expect that,
once stringent selection criteria are applied to events, 
distinctive features 
associated to a HS would become useful in the search strategy
and subsequent interpretation of new phenomena
in multiparticle dynamics.

\subsection{Hidden sectors}
In this work we focus on Unparticle
physics \cite{Georgi:2007ek}
in particular, though the main ideas can 
be generally applied to Hidden Valley models \cite{Strassler:2006im}. 
In these models, the Standard Model (SM) is accompanied by
a HS of new particles not been yet observed due to typically
an energetic barrier or a weak coupling to SM particles.
We will assume further that particles coming from a
HS can decay back to SM particles \cite{Strassler:2006im}, 
thereby modifying the 
conventional pattern of
the parton cascade in multiparticle production.

So far most signatures of new physics (like jets, missing energy, high-$p_T$
leptons or photons, displaced vertices) have been considered 
on the transverse plane with respect to the beam direction. 
In a complementary way, we advocate that a new
stage of matter might also show up in {\em soft} physics of 
underlying events {\em tagged by hard products}, e.g. through
particle (pseudo)-rapidity correlations (either integrated or not).

\begin{figure}
  \includegraphics[height=.2\textheight]{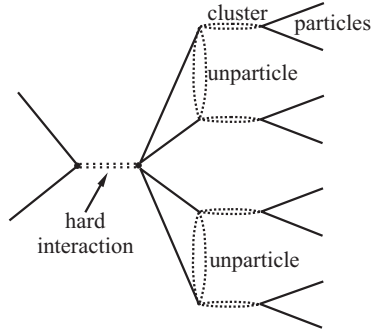}
  \caption{Pictorial representation of a 3-step
scenario where unparticles (for our particular HS choice)
are produced at a hard parton interaction in $pp$ collisions, 
subsequently decaying into
final-state SM particles through cluster formation.
The number of unparticles sources
at the onset of the cascade may fluctuate,
and a large (and $\lq\lq$continuously variable'') mass
of the unparticle stuff would induce additional 
long-range correlations among the
final-state SM particles.}
\end{figure}

\section{Inclusive (longitudinal) rapidity correlations}

Experimental results in multibody production along decades
have steadily supported the tendency of produced particles to merge
into correlated groups \cite{Kittel:2005fu}. 
This experimental evidence has traditionally
led to the view of a two-step process for high-energy hadron 
collisions. The resulting multiplicity distribution
is thus given by the convolution for particle emission sources
(strings, clusters/clans, fireballs...) with
the decay/fragmenting distribution of sources. 
However, observable consequences 
could derive from a new stage of matter 
(stemming from a HS) generated at the onset
of the parton shower (see Fig.1)  
leading to a 3-step scenario in hadronic
collisions at very high energy.

The inclusive 2-particle (rapidity) correlation function is defined as
\begin{equation}\label{c2}
C_2(y_1,y_2)=\frac{1}{\sigma_{in}}\frac{d^2\sigma}{dy_1dy_2}-
\frac{1}{\sigma_{in}}\frac{d\sigma}{dy_1}\frac{d\sigma}{dy_2}
\end{equation}
where $\sigma_{in}$ denotes the inelastic $pp$ cross section
and subscripts 1 and 2 refer to the two considered particles
event by event. $C_2(y_1,y_2)$ is usually split in two terms:
\begin{equation}
C_2(y_1,y_2)=C_2^{SR}(y_1,y_2)+C_2^{LR}(y_1,y_2)
\end{equation}
where the short-range (SR) part is generally assumed to be more sensitive 
to dynamical correlations, while $C_2^{LR}(y_1,y_2)$ stands for
long-range (LR) correlations usually due to the mixing of different
topologies. In this work we focus on the central rapidity region.

According to our study \cite{SanchisLozano:2008te}, the 
$C_2^{LR}(y_1,y_2)$ piece in Eq.(2) should become enhanced in a 3-step
scenario wrt a standard 2-step cascade as a consequence
of larger fluctuations of the number (and mass) of
the primary sources of partons. In addition, 
the correlation length in the $C_2^{SR}(y_1,y_2)$ term should become 
larger because of the (presumably) higher mass of the intermediate 
hidden (un)particle stuff. 
Therefore longer and stronger correlations 
between particles can be expected from both SR and LR terms
in Eq.(2) whenever a HS appears in the cascade. Nevertheless,
other conventional sources of possible LR effects in
multiparticle production \cite{Ryskin:2009qf}
should be properly taken into account.

\begin{figure}[ht!]

\includegraphics[scale=0.5]{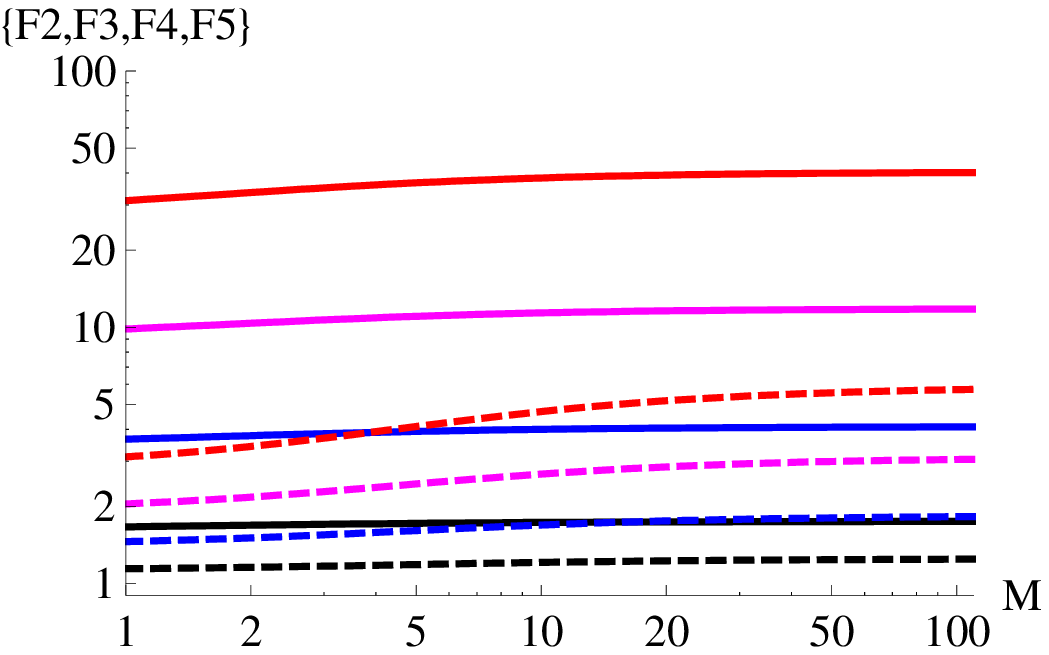}
\includegraphics[scale=0.45]{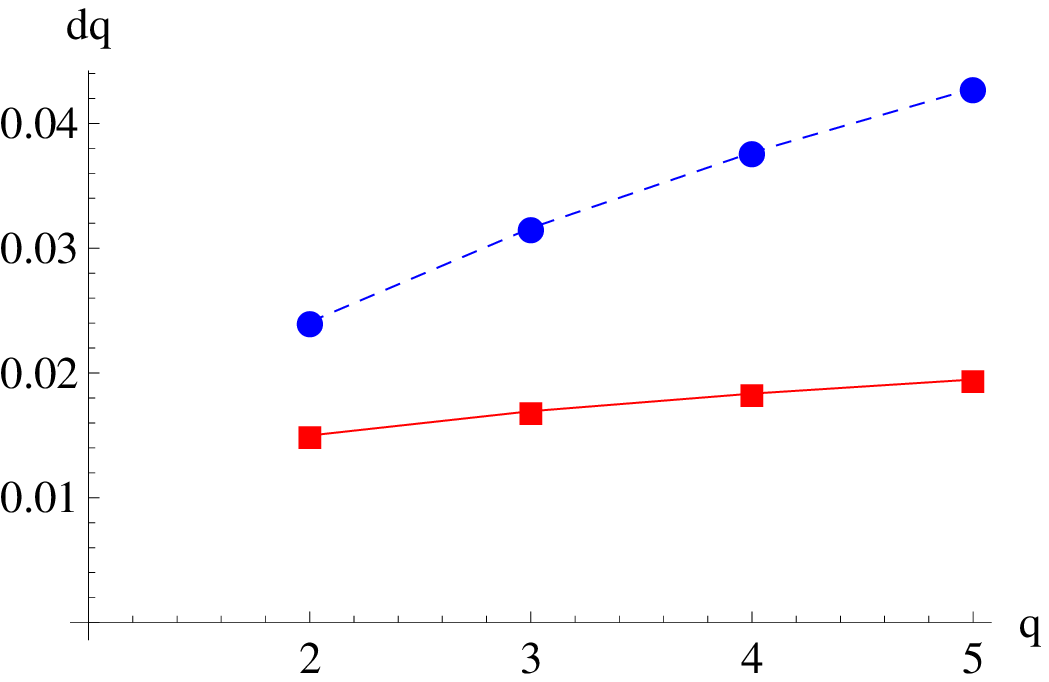}
\includegraphics[scale=0.45]{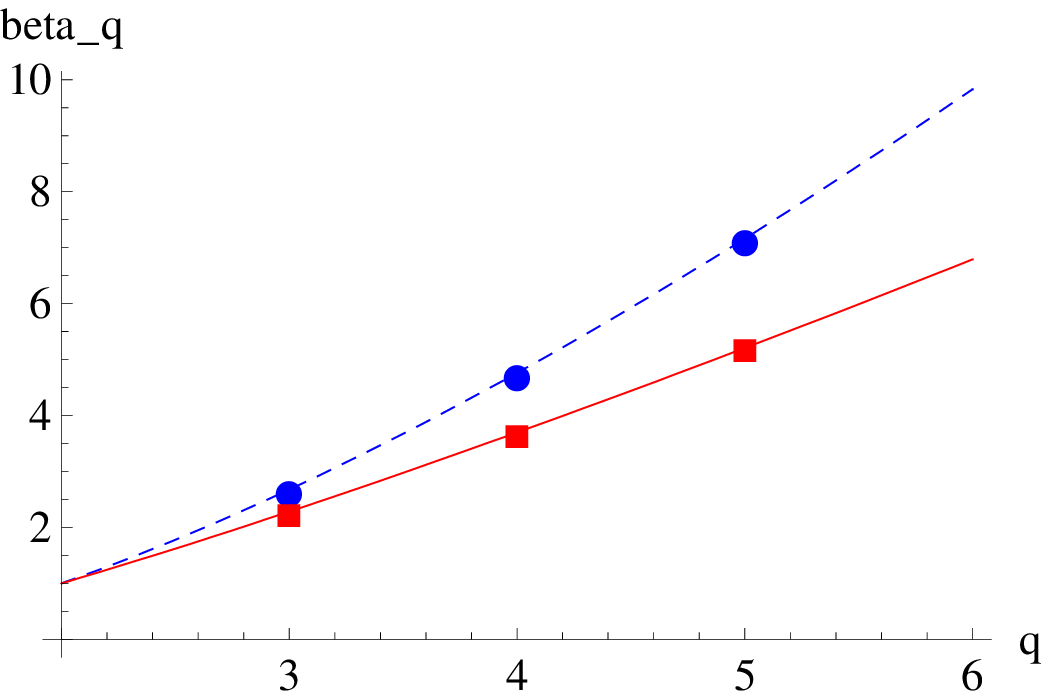}
\caption{Left: $M$-scaling for $q=2,3,4,5$ (upwards), in 2-step  
(dotted lines) and 3-step (solid lines) cascades;  
Middle: $d_q$ versus $q$ for 2-step and 3-step cascades;  
Right: $\beta_q$ versus $q$ setting $\mu=1.46$, $\nu=1.42$ 
($\mu=0.72$, $\nu=1.18$) in a 2-step (3-step) cascade,
respectively. See \cite{SanchisLozano:2008te} 
for more details.} 
\label{fig:Dq}
\end{figure}

\subsubsection{Intermittency, multifractality and entropy}

Fluctuations in small phase space regions (intermittency)
have been commonly described by the scaled moments $F_q(\delta y)$
of a multiplicity distribution $P_n$ 
(corresponding to the normalized phase-space integral over
the $q$-particle density function) as 
the rapidity interval under study $\Delta Y$
is split into $M$ bins of equal size 
$\delta y=\Delta Y/M$ \cite{Bialas:1985jb}. 

On the other hand, the fractality nature
of multiparticle hadroproduction 
is deeply connected
with intermittent behaviour exhibiting a power-law
dependence of the multiplicity moments with the cell size 
\cite{Bialas:1985jb}. 
If self-similar dynamical fluctuations exist, $F_q(\delta y)$ should
obey a power-law increase ($M$-scaling) at small $\delta y$, i.e.
\begin{equation}{\label{eq:powerlaw}}
F_q(\delta y)\ \sim\ \delta y^{-\phi_q}\ \sim M^{\phi_q}\ \ 
\to\ \ \ln{F_q}=a_q - \phi_q \ln{\delta y} \equiv A_q+\phi_q \ln{M}
\label{eq:loglaw}
\end{equation}
where the $\lq\lq$intermittency exponents''  $\phi_q$ ($\phi_q > 0$)
are related to the anomalous fractal dimensions 
$d_q$ \cite{Lipa:1989yh,Hwa:1989vn} as 
\begin{equation}
d_q=\frac{\phi_q}{q-1}
\end{equation}

Moreover, the power scaling between 
$F_q$ and $F_2$ ($F$-scaling),  
\begin{equation}\label{eq:Fscaling}
F_q(M) \sim [F_2(M)]^{\beta_q}\ \ \to\ \ 
\ln{F_q}(M)=\beta_q\ln{F_2}(M)
\end{equation}
represents another interesting relation, 
where the $\beta_q$ coefficient can be expressed as
\begin{equation}\label{eq:beta} 
\beta_q=\frac{\phi_q}{\phi_2}=\frac{d_q}{d_2}\dot (q-1)
\end{equation}
Monofractality ($d_q=d_2$, $\forall q$) implies that $\beta_q=q-1$,
which can be tested from (5).

Brax and Peschanski \cite{Brax:1990jv} proposed
a better approximation than  Eq.(\ref{eq:beta})
using a L\'evy stable law description of multiparticle
production:
\begin{equation}\label{eq:Levy}
\beta_q=\frac{\phi_q}{\phi_2}=\frac{q^{\mu}-q}{2^{\mu}-2}
\end{equation}
where $\mu$ is the L\'evy index (also known as the
degree of multifractality) which permits an estimation
of the cascading rate; $\mu$ should be
in principle restricted to the interval $0 < \mu \leq 2$ 
(region of stability) with $\mu=0$ 
characterizing monofractal behaviour.

Instead of the law expressed in Eq.(\ref{eq:Levy}), $\beta_q$
has been parametrized in the Ginzburg-Landau model of phase transitions
as  
\begin{equation}\label{eq:nu}
\beta_q=(q-1)^{\nu}
\end{equation} 
where $\nu=1$ now implies monofractality. 
The motivation for all these approaches 
in heavy-ion collisions comes from the need of
a signal for Quark-Gluon Plasma formation through, e.g., 
a (second-order) phase-transition, since the correlation
length would then diverge, and the system behave as a simple fractal 
\cite{Bialas:1990xd}.

We are certainly not considering here such kind of phase-transition, but 
extra LR correlations from a HS leading to a similar effect 
on both $\mu$ and $\nu$ estimators.
Indeed our results shown in Fig.2 are consistent
with the interpretation that multiparticle
production including a HS approaches monofractality; the 
intermittency exponents, all $d_q$ values and $\mu$ and $\nu$ estimators 
turn out to be smaller in a 3-step scenario than in a 2-step cascade.

Let us finally note that the (Shannon) entropy,  
defined for a multiplicity distribution $P_n$
as $S=\sum_{n}P_n\ln{P_n}$ \cite{Simak:1988qp}, 
might become another indicator of a HS  
in multiparticle production. For example, it is relevant
to verify to what extent entropy is additive 
\cite{Bialas:2006sy} by checking  
whether $S(R)=S(R_1)+S(R_2)$ using
particles belonging to regions $R_1$ and $R_2$ 
well separated in rapidity space
(e.g. forward and backward hemispheres). 

\section{Summary}

Intermittency and (multi)fractality 
can be sensitive to a HS altering 
the pattern of the parton cascade in high-energy inelastic
$pp$ collisions. From our study in \cite{SanchisLozano:2008te} 
we conclude that longer and stronger correlations between
emitted particles, and larger scaled moments in multiplicity distributions,
should be expected than in a conventional QCD cascade, while
the pattern of the multiparticle system approaches
monofractality - or merely becomes less fractal. The
challenge remains however in tagging those {\em anomalous} underlying 
events (where such new phenomena would show up through soft physics)
by appropriate selection cuts, to be compared with
a control sample of standard events. More
work requiring prior tuning and Monte Carlo generation 
of events at LHC energies
is required before drawing any conclusion about the  
feasibility of our proposal.



\begin{thebibliography}{9}

\bibitem{Kittel:2005fu}
  W.~Kittel and E.~A.~De Wolf,
  {\em Soft Multihadron Dynamics}, edited by
World Scientific, 2005.


\bibitem{SanchisLozano:2008te}
  M.~A.~Sanchis-Lozano,
  arXiv:0812.2397 [hep-ph]. 


\bibitem{Georgi:2007ek}
  H.~Georgi,
  Phys.\ Rev.\ Lett.\  {\bf 98}, 221601 (2007)
  [arXiv:hep-ph/0703260].


\bibitem{Strassler:2006im}
  M.~J.~Strassler and K.~M.~Zurek,
  Phys.\ Lett.\  B {\bf 651}, 374 (2007)
  [arXiv:hep-ph/0604261].


\bibitem{Ryskin:2009qf}
  M.~G.~Ryskin, A.~D.~Martin, V.~A.~Khoze and A.~G.~Shuvaev,
  arXiv:0907.1374 [hep-ph].


\bibitem{Lipa:1989yh}
  P.~Lipa and B.~Buschbeck,
  Phys.\ Lett.\  B {\bf 223}, 465 (1989).


\bibitem{Hwa:1989vn}
  R.~C.~Hwa,
  Phys.\ Rev.\  D {\bf 41} (1990) 1456.



\bibitem{Bialas:1985jb}
  A.~Bialas and R.~B.~Peschanski,
  Nucl.\ Phys.\  B {\bf 273}, 703 (1986).


\bibitem{Brax:1990jv}
  P.~Brax and R.~B.~Peschanski,
  Phys.\ Lett.\  B {\bf 253}, 225 (1991).


\bibitem{Bialas:1990xd}
  A.~Bialas and R.~C.~Hwa,
  Phys.\ Lett.\  B {\bf 253}, 436 (1991).

\bibitem{Simak:1988qp}
  V.~Simak, M.~Sumbera and I.~Zborovsky,
  Phys.\ Lett.\  B {\bf 206}, 159 (1988).

\bibitem{Bialas:2006sy}
  A.~Bialas, W.~Czyz and K.~Zalewski,
  Acta Phys.\ Polon.\  B {\bf 37}, 2713 (2006)
  [arXiv:hep-ph/0607082].



\end{thebibliography}
\end{document}